# Inattentive in social, active in mind: VR-based design intervention for imagining desirable possibilities in the public space

Yiying WU*[a], Miikka J. Lehtonen [b]

[a] School of Design, The Hong Kong Polytechnic University, Hong Kong SAR, China

[b] College of Business, Rikkyo University, Tokyo, Japan

* bow.yiying.wu@gmail.com

The metro as a form of public transportation is an important urban infrastructure that takes a large population from place A to B every day. To achieve that, it is primarily designed for extreme functionality and efficiency. However, in terms of experiential aesthetics, the metro is seldom people's favourite place. When this modern infrastructure succeeds in serving urban mobility with high performance and efficiency, passengers seem to want more than the guaranteed functional performance. Recently, with the emergence of Virtual Reality (VR) technologies, increasing efforts from design and HCI communities look at the value of VR technology in enhancing commuting experiences, bringing new possibilities of interaction and activities, and potentially transforming social public spaces. This study investigates how and why VR technology could be integrated with a metro ride. We experimented with ten passengers by showing them three 360° videos during their metro ride. The results show the narrative-driven scene is most desirable. Despite wearing a VR headset might cause anxiety, our findings indicate a high level of acceptance towards VR experiences based on the finding that it does not challenge the normative behaviours of being a passenger 'inattentive in social, active in mind' and further can enhance the experience. As the takeaway, we propose three strategies of VR content tailored for the metro context in which passengers would find a role participating in the virtual scene and turn the scene to one's own story, and at the same time, maintain physically constrained.

## 1 Introduction

The metro as a form of public transportation is an important urban infrastructure that takes a large population from place A to B every day. To achieve that, it is primarily designed for extreme functionality and efficiency. However, in terms of experiential aesthetics, the metro is seldom people's favourite place. It is often related to the dystopian vision of boredom and dullness. Metro stations and cabins are grey and monotonous, and passengers appear passive and constrained and also indifferent to each other. When this modern infrastructure succeeds in serving urban mobility with high performance and efficiency, passengers seem to want more than the guaranteed functional performance. Responding to this need, designers spiritualize public spaces by creating more pleasant, artistic or convivial experiences (Mattelmäki & Vaajakallio 2012, Remesar 2021) by various strategies of the community (Remesar 2021) or human movements (Vroman & Lagrange 2017). Recently, with the emergence of interactive technologies, increasing efforts look at the value of technologies such as sensor-based devices (Kuznetsov & Paulos 2010), public screens (Seeburger & Foth 2012) in enhancing commuting experiences, bringing new possibilities of interaction and activities, and potentially transforming social public spaces. More recently, McGill et al. (2019) explored the use of immersive Virtual Reality (VR) technologies in assisting entertainment or work while people are commuting.

This paper aims to explore the possibility of VR technologies in bringing new commuting experiences and activities to the metro. VR applications are defined as either 3D VR or 360° video where the former is usually created in digital environments whereas the latter is shot in actual settings (as per Yang et al. 2021). Moreover, 3D VR is interactive while 360° videos are immersive. While both applications are experienced with VR headsets, on an experiential level they differ quite dramatically. Throughout the course of this paper, we focus on 360° videos. The study draws on a pilot study in the capital region of Finland with the method of field-based design intervention. In this study, we created three rudimentary 360° videos focusing on nature in the vicinity of the metro itself and showed it to ten participants on a standalone VR headset during their metro ride.

Our enquiry includes two parts. First, we examine the social acceptance of passengers using VR technology in the metro ride. We mainly approach the acceptance by analysing the relation between the intervention and normative behaviours in the commuting context. By analysing what norms had and had not been challenged by the intervention, we generate insights of acceptance in terms of social behaviours. Second, we investigate what types of immersive 360° video content are more desirable or suitable in the commuting context. Suggestions related to VR content will be provided. The pilot study wishes to contribute to the discussion on and the possibilities of developing VR technology as part of metro services and VR experiences part of daily commuting experiences, and further the potential of VR technology in transforming social and public spaces.

Our work mainly draws on two main bodies of work: the theoretical study of urban behaviours from urban sociology (Goffman 2005, Simmel 2012) and the study of wider applications of mobile and nomadic VR technology in public spaces (Gugenheimer et al. 2019, Eghbali et al. 2019). Firstly, the urban theories provide a different view to the observed behaviours in the metro and related public spaces that are often been criticised as dull, passive and indifferent. They argue that passivity and indifferences are the socio-spatial reality, and the two characteristics have made the urban space functional and furthermore enable another form of liveness and imagination (Goffman 2005, Mäenpää 2012). This fresh account helps us to analyse the empirical data of participants' routinized metro-related experiences and activities and the intervention result. Secondly, the current work from HCI and interaction design that study the application of VR technology in the wild mostly focus on investigating social acceptance. Some key obstacles are commonly identified such as lack of reality awareness, detachment from other people nearby, or security (Gugenheimer 2016). Also, acceptance is largely examined from the aspect of user experiences. For instance, Eghbali et al. (2019) identify four experience categories of autonomy, security, popularity and relatedness related to the social acceptance of VR applications in restaurants. Our study shares the affinity with the previous study of social acceptance from the aspect of experiential factors and takes a step further to investigate the desirability of video content in the specific context of the metro ride.

## 2    Related work

We firstly present two different views of 'A dull and passive urban man' and 'a happy urban man' of the metro and social behaviours. Then we review the work from the fields of HCI and interaction design that investigate main challenges and obstacles, design implications and solutions when taking VR technology to the wild.

## 2.1 A critical view: A dull and passive urban man

The public space of the metro is viewed critically in cultural discourse that the metro is dull and over-regulated and users are passive and anonymous (eg., Edensor 2011). Popular cultural discourses describe 'commuting as a dystopian, alienating practice, and the commuter as a frustrated, passive and bored figure' (Edensor 2011). Augé (1995) calls the metro one of the 'non-places'. Opposite to 'a place' where people would like to stay, the metro, among other infrastructures of transport of high-speed roads, railways and train stations, is built to shift passengers in an extremely fast and efficient manner. Thus, the quality of such places is measured by whether people can pass through with the fastest speed. As a result, it does not hold social relations or cultural and historical identity. On the one hand, while waiting to be shifted away, people seldom develop interesting activities with the surrounding and meaningful interaction with each other. Thus, they remain anonymous and isolated (Cresswell 2006). On the other hand, the metro is primarily governed by rules and dominated by big business in search of profits (Lefebvre & Nicholson-Smith 1991). Hardly metro space has the opportunity to express the voices and needs of local communities or commuters. And the regulation and control as well increase the passivity of people while using the public space.

In all, the metro, like many industrial urban places, is the nightmare for the researchers who pursue socialization, localization, and artistic quality and experiences in public places. As responses to the above-mentioned problems, designers and artists attempt to use the interventionist approach to turn the metro to more lively places. They believe that the metro has the potential to stage socio-cultural experiences and activate people. Among many cases, we take a two-year Project 'Spice' from Helsinki as an example that intended to spiritualize and enrich public places in the metro (Mattelmäki & Vaajakallio 2012). One young designer explicitly expressed his dissatisfaction with the scene of an urban man spending another gloomy day in the metro station and showing forced loneliness together due to the behavioural code of keeping distance. One design concept was to project passengers' photos recorded by surveillance cameras in real-time to the big screens near the platform (Kola 2011).

## 2.2 A different view: A happy urban man

However, the behavioural codes of anonymity and passivity are seen differently by others. Simmel (2012) argues that anonymity that has made a great modern city gives freedom to each individual to act, express and present oneself in one's own way without being afraid of being judged by the crowd. Thus, anonymity in urban public space is an effective social force. And relating to indifference, Goffman (2005) calls it 'civil inattention'. It means when we claim a person appears indifferent to others in public space, the more accurate interpretation shall be s/he is acting actively to show 'I am not attentive to you' to others. If not showing the inattention explicitly, it would be considered as rude and an invasion of other's private territory in many social situations. In our study of Finnish metro experiences, many subjects mentioned the unpleasant moments of being approached by drunk people who showed interest in talking to them. And they had to act hard to show indifferences and ignore the interactive invite. Another example is in the crowded elevator. When there is little space left and more chances to have eye contact among people, everybody starts to avoid giving attention to others, either by looking at the roof, one's own shoes or the phone in a very forceful and dramatic manner.

Next, we look at the second term of passivity by giving a different perspective 'passive in body, active in mind'. Mäenpää (2012) argues that when an individual appears dull and lonely, one might be actually active and busy in the inner mind in which imagination prospers. The urban man wanders around the urban space that is composed of giant advertisement boards, well-decorated windows of shops, cafes, restaurants and infrastructures. They are busy scanning, interpreting and absorbing various and intensive visual and experiential elements from advertisements for instance. These are inspirational resources for one to project and imagine an ideal or refreshing self in an imagined scenario. This is the 'self-illusory and hedonistic' man in Simmel's sense (2012). Similarly, as Sennett (2017) points out, a man in the public space is no longer acting for or with other people but more for oneself. This is where pleasure comes from, which is through the constant imagination of an interesting self. In this way, other strangers are not for talking with but used as resources to feed one's imagination.

The perceptions stated above provide us with a deeper perspective to analyse and understand the socio-spatial reality in metro cabins and the normative behavioural codes of passengers. They reveal the mechanism that the urban space is composed of strangers who give each other inattention, respect and freedom of expression, and at the same time are busy diving into the construction of the imagined self.

### 2.3 VR and 360° videos in the wild

VR applications have been experimented with in various types of public places like cafes, restaurants (Mai & Khamis 2018, Schwind et al. 2018) as well as in cars (Paredes et al. 2018, Hock et al. 2017) and aeroplanes (Williamson et al. 2019). And 360° videos have often been studied in the context of tourism (Rahimizhian et al. 2020, Yang et al. 2021). Due to the relative novelty of VR technology, there is a growing body of work focusing on acceptance and resistance issues to contribute to wider applications of VR technology. The resistance is from some widely recognised worries of losing connection with people nearby (Mai et al. 2017), losing awareness from real surroundings (O'Hagan & Williamson 2020), and security. These obstacles are brought by the inherent feature of VR 'immersion' (Kelling et al. 2017). The HMDs that cover the wearer's eyes make one's face and especially eyes unseen by others, thus drastically causing the communication barrier between wearers and bystanders and decreasing awareness of the happenings around (McGill et al. 2015, O'Hagan & Williamson 2020). Responding to these obstacles, researchers add reality awareness functions such as notifying bystanders' presence or sensitising the reaction to audio happenings (O'Hagan & Williamson 2020). Some studies compensate for the lack of direct interpersonal interaction by making HMDs 'transparent' (Mai et al. 2017, Chan & Minamizawa 2017, Yang et al. 2018, Gugenheimer et al. 2017). Moreover, in order to facilitate the interaction between HMDs wearers, various interaction techniques are proposed such as smart watch (Hirzle et al. 2018), the back of the HMD (Gugenheimer et al. 2016), and magnet-related (Smus & Riederer 2015).

In the context of travel, VR passenger experiences have been examined (McGill & Brewster 2019, McGill et al. 2019). There are several practical reasons concerning social acceptance, like safety, awareness of other passengers, notification of getting off (Gugenheimer et al. 2019) and motion sickness (Akiduki et al. 2003). Many researchers have worked on solutions to reduce the negative experiences caused by motion (McGill et al. 2017, Hock et al. 2017), such as creating the visual element of VR content aligned with the real car movement (Paredes et al. 2018, Akiduki et al. 2003), and mapping vehicular movements with visual information of VR content (Hock et al. 2017). Also, it is important to keep a good balance between immersion with the virtual world and connectedness with the real world (Williamson et al. 2019, Ahmadpour et al. 2016). The interruption which is often conveyed through voice or touch needs to be carefully reconsidered because a light touch can be alarming during VR experiences (Groening 2016). New functions shall facilitate and support these needs, especially providing peripheral awareness to wearers (Williamson et al. 2019).

## 3 The case and method

### 3.1 Field-based design intervention

This pilot study is informed by the method of field-based design intervention to approach the enquiry of the potential of developing VR technology in transforming public spaces. Halse & Boffi (2016) regard design intervention as a research method to enable new forms of experience and practice to emerge around the subject and understand and explore issues, concerns and dialogues around the design space. It draws on sources from social science like 'creative disruption of everyday life' (Sholette & Thompson 2004). It is an exploratory research method that combines qualitative empirical research and the generative methods of constructive design research. This method is especially useful in the early stages of design exploration in which things are yet to be coherent, settled, or fully articulated (Halse & Boffi 2016). Design prototypes or mock-ups are made and brought to real social situations for people to use. When intervention is introduced into the field, the relations among variables within the system are studied. Our process of data analysis is implied by the analytic framework 'Ethnography of design intervention' in understanding the concrete practices, behaviours, emotions, and experiences emerging from design interventions (Wu 2017). This framework holds dual positions by answering 'what is challenged' and 'what is emerging'. The first looks at the existing structural elements in the intervened setting. By analysing what norms had (not) been challenged or reinforced by the intervention, we generate the insights of acceptance in terms of social behaviours. Only if the technology does not subvert existing norms and further is able to enhance existing activities, there is more chance for the new technology to be accepted (Davis et al. 1989). And the second 'what is emerging' is speculative, studying the new, the possible and emerging. By looking at what new content and experiences are more desired by participants, we propose new experiences that VR applications would bring to metro rides. Both dimensions are crucial for our studies. The reflective position helps investigate the social acceptance of VR applications in the context of the Finnish metro, and the speculative one helps explore what new 360° content to bring to passengers.

It is worthy of note here that the role of the three 360° videos is closer to technology probes (Hutchinson et al. 2003) or design probes (Mattelmäki 2006). They are not meant for evaluation or testing. Rather, used as a probe, they are used for design concept development, through which we learn about passengers' preferences, provoke imagination and thoughts from passengers, and identify further design focus or direction. Therefore, by adopting this approach with more design- rather than evaluation-oriented, the sample of ten participants is considered as appropriate.

## 3.2   Case: Bringing nature to metro

In November 2017, the Helsinki Metro extension was opened, thus including Espoo, a neighbouring city, in the network. Consequently, while most passengers use the metro to commute, people have also utilized the new metro extension to visit various areas in the capital region. In this context, we collaborated with officials from the City of Espoo to explore whether VR applications in the metro could not only improve and enrich commuting experiences, but also provide a good opportunity to show and promote local beautiful nature scenes to passengers. We regarded that the brief of nature from our collaborator was an appropriate topic as introducing 'nature' to the metro would provide a reasonable distance from reality. Although the local nature scene was about the experience of the familiar and mundane instead of the exotic far away, it provided experiences of plus and otherness beyond the usage context of the metro. By providing local scenes, we also wished passengers to associate the elements from local scenes with their personal experiences, stories and memories of nature and further strengthen the local identity.

At the same time, we were interested in the complex relationship between nature and urban life. As the rate of urbanization and utilization of technologies in everyday life has been steadily increasing during the last few decades, we witness a certain longing for nature and experiences in nature (Turner et al. 2004). Metro represents fast speed, urban space, modern technological infrastructure, and stressed work life, while nature represents slowness, pause, and relaxed leisure time. What would be the possibilities of using VR applications to mediate the two directions of experiences? While we are not advocating virtual environments and virtual reality to offer a reconciling path between urbanization and alienation from nature, we do ask whether the combination of nature and technologies could offer evocative experiences in non-spaces and liminal spaces such as metro cabins.

Thus, we created three rudimentary 360° videos of nature scenes including distinctive Finnish features of lake, forest, summer cottage, bone fire, grill, ocean, fishing and boat (Figure 1). The first video 'Lake' presented a typical Finnish rural summer cottage. It is in general static, composed of good sunlight, forests, lake and a cottage. However, it is active in details like bird singing, butterfly flying, leaves rustling with the breeze and the rippling sound of the lake. The second portrayed a man making bonfires and grilling on a dark night. The third was about two men going for a fishing trip with a motorboat to the open sea. For the experiment, the videos were presented to the participants back-to-back in the aforementioned order.

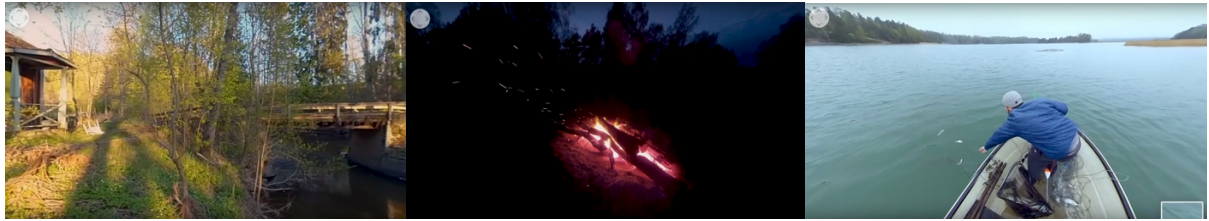

*Figure 1. Video 1 (left): 'Lake', one minute long; Video 2 (middle): 'Bone fire', two minutes, Video 3 (right): 'A fishing trip', four minutes. Copyright: Johannes Söderström.*

### 3.3 Data collection

Data for this study was gathered through participant observation of both participants and standby passengers and one-to-one interviews with ten participants in the 2019 summer. We showed the three videos lasting seven minutes in total to each participant on a VR headset during their metro ride (Figure 2). We used the standalone VR headset 'Oculus Go' because this model had an inbuilt screen and memory with no request for Wifi.

The ten participants had balanced portions in gender (male: N5; female: N5) and age (18-29YO: N5; 30-39YO: N3; 40-49YO: N2). All live in the capital area of Finland and use public transport as the main means of daily commuting. Of the ten, six are Finns (three were born and grew up in Helsinki, and three moved from smaller cities) and four are of diverse nationalities. Three had never experienced VR before our intervention, while seven had some experiences with VR technology, including one owns a VR headset to play games at home. All expressed their positive opinions of VR technology which could enable them to experience something different or new from ordinary reality. However, none had experienced VR in the public before our intervention.

Pre- and post-intervention interviews were conducted. In the pre-interview, the focus was on existing practices of daily metro commuting and memorable experiences relating to nature. The post-interview was conducted immediately after the intervention at places near the metro station. The focus was on sensations, emotions, and presence with regards to the intervention to better understand how the participants made sense of these experiences. Before and after the invention, each participant was asked to rank their commuting experiences with a scale from +2 (strongly agree) to -2 (strongly disagree). The eight attributes (Figure 3) were developed from the previous VR work designing for experiences (Kauhanen et al. 2017).

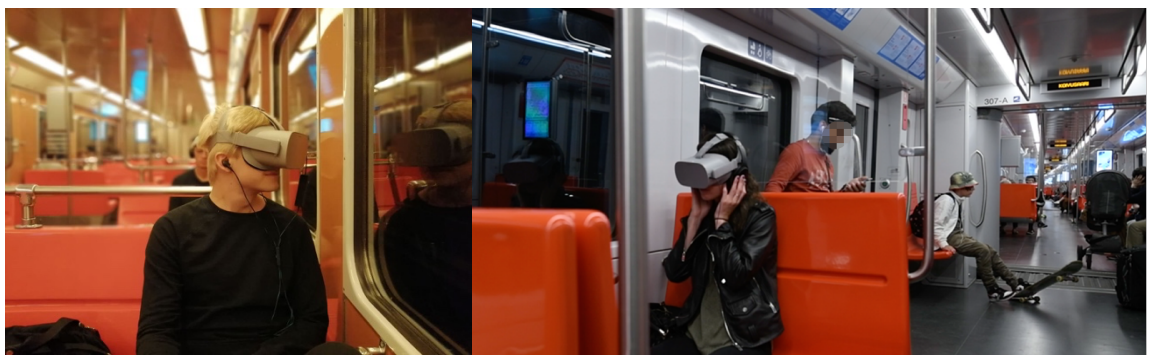

*Figure 2. Participants were experiencing 360° videos during their metro rides. Copyright: Johannes Söderström.*

## 4 Results

Overall, most participants (except P5) regarded their metro experiences as satisfactory, except for the stressful part which was mainly attributed to crowded cabins in rush hours. This result shares similar findings from the studies of railway commuting in Japan at present (Negishi & Bissell 2020) and of London subway a decade ago that reveals that 62% of commuters found ways to keep themselves busy and only 2% found travelling boring (Lyons et al. 2007). From our result, all agreed the Finnish metro was a very safe place, and most chose 'relaxed', 'not bored', 'calm' and 'time flies fast' to describe their metro trips (Figure 3). Although the metro cabin itself was mentioned as visually and experientially boring the same as perceived by the designers (Mattelmäki & Vaajakallio 2012), each participant found various satisfactory ways (eg., reading books, listening to music or podcast, texting friends) to fill their travel time and cope with boredom.

After trying the VR headset for seven minutes during the ride, except P7 who reported light dizziness and motion sickness, all participants commented it as interesting and comfortable. Moreover, the intervention enhanced sensations of calmness, relaxation, and reduced the negative ones of boredom and stress. However, concerning the negative changes, security was the main cause of concern. The intervention dramatically reduced the feeling of security from score 18 to 7 (Figure 3). However, in this Finnish case most still considered the intervention as 'safe' albeit with the worry expressed. We think this evaluation result might be because they were aware of this experiment and there was an observer whom they could trust.

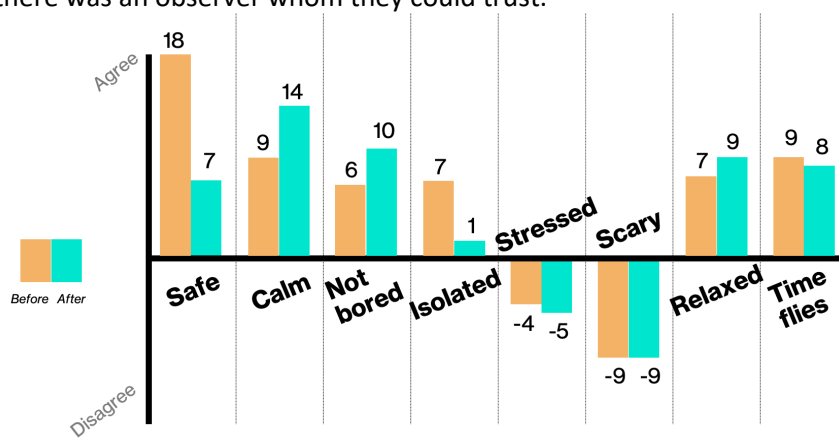

*Figure 3. Before (orange): the result of participants' usual commuting experiences of the metro; After (green): the result of the new experiences with the VR intervention. The score is the total sum of the scores from ten participants.*

The second part describes the discussion on preferences, experiences and wishes provoked from watching the three videos. The preference from ten participants was consensual. As a summary, Video 3 was participants' favourite which was considered immersive, fun to watch, while the other two did not bring surprising or impressive experiences. Video 1 'Lake' triggered the experience of boredom more than calm. Although in the pre-interview many participants mentioned they loved visiting peaceful lakes in reality, they did not associate peace or feeling of mediation with the lake in the 360° scene. Only three participants considered themselves 'calm' or 'relaxing' while watching this clip *('I can imagine I can put my hands behind my head and start to look around; feel like I can stay here as long as I want'*), while the rest found it boring as 'nothing is happening' (P5). With deep boredom, P2 mentioned in the interview that she was desperately waiting for something to happen like a deer running into or a guy walking out of the cottage. And the second video 'Bone fire' was everyone's least favourite mainly due to the darkness and poor resolution that participants could barely see anything.

Video 3 'A fishing trip' was almost everyone's favourite except P5 who was bothered by the fact that the 360° camera was not positioned at the right place. There were four main elements that made this clip most entertaining for participants to watch. The first was the movement. The sensation of the movement of the boat in the video synchronized with metro movement, which added to the immersive feeling (P1: *'Since the boat moved and the metro did, I felt really natural and being immersed.'*). Also, while watching, two participants were adjusting their sitting position as if they were leaning on the boat and attempted to keep the balance. Secondly, this video had a storyline, with the start of 'taking off to the ocean to catch fish' and ended with the result of 'we caught fish'. Along the process, there were new things taking place continuously. Overall, a story with a thread made the audience easy to follow with constant curiosity. The third element was the immersive companion. As P2 remarked, having other two people in the scene made him feel 'I am the third person on the boat' that the two previous videos did not give. And this added to his experience related to active immersion and participation. Fourthly, there were a variety of elements to see: the scene composed of open sea, seagull, the movement; the activity that required skills and usage of different kinds of equipment; and human companions. Participants could find one's own interest from the single video. For instance, it brought P4 back to her childhood memory of going to the sea. P6 focused on the fishing practice itself and considered it a good chance to learn a new skill. P2 had fishing experience himself and closely inspected the tools used. P1 simply enjoyed the movement of the boat.

## 5   Findings

This section presents two main findings from this study. In answering 'what is challenged' to the enquiry of social acceptance, we argue that the introduction of VR technology to metro space and commuting practice conforms to the socio-spatial reality and does not challenge much or alter the existing normative codes. And it further shows VR can provide better experiences to a small extent for passengers who already have overall positive commuting experiences.

### 5.1   VR technology enhancing 'passive in body, active in mind'

In terms of social behaviours in the metro context, there is a high level of acceptance of applying VR technology during the metro ride. This conclusion is based on the finding that the new practice of passengers watching 360° videos with VR headsets did not fundamentally challenge or subvert the normative codes of behaviours in the Finnish metro. Furthermore, VR technology-enhanced commuting experiences in the same way to the means currently that were applied by participants.

Firstly, the new practice of watching with VR headsets fit the existing pattern of inattention and detachment. As observed from the empirical study, the visual appearance of wearers wearing the headset was recognisably different from others, but their behaviour remained almost unchanged. With the VR headset, one sat still and passive on the seat. What had been challenged speaking of norms? The new technology made the wearer's visual appearance recognisably distinctive and, at the same time, reinforced the detachment. We argue both would receive a high level of acceptance. Regarding the first norm, being visually distinctive is not unusual in a metropolis where people have a higher chance to encounter others wearing the newest digital gadgets or sub-cultural outfits. Regarding the second, the empirical data shows we got much less attention from other passengers in the field than we expected. Partially the reason is half of the time the metro cabin was rather empty which means no passenger sat around the participant. However, there was one time when a girl was sitting next to the participant, she did not pay attention to the participant throughout the process. Although we did not inquire into the opinions of standby passengers, at least they acted in the way in which they delivered inattention to participants in order not to be rude. As Goffman (2005) argues, showing 'civil inattention' to other anonymous people in the urban public is the basic norm. We suggest the increased level of detachment brought by VR headsets will not reduce the level of acceptance as it does not break the balanced socio-spatial structure of the metro cabin. Contrarily, if 360º videos are designed to encourage passengers to behave actively or develop interaction with the surrounding, such intervention would cause inconvenience and thus be less accepted. Therefore, appropriate VR content needs to be carefully designed in order not to challenge the passive behavioural norms.

Secondly, while passive in body, VR technology enhances the status of active in mind. As reported from the interview, the shared pattern of their activities in the metro was they created and indulged in one's own bubble world and exaggerated inattention to and detachment from the surrounding. They often wore headphones, closed eyes, lowered their gaze to stare at the phone, or chose to sit in the corner. In the personal bubble built temporarily by music, video or book, they were seeking joy and inspiration. In this sense, wearing VR headsets is the advanced edition of the means of creating one's own bubble world and seeking joy from another world.

It provides implication to the broader enquiry of designing for immersive digital technologies such as smartphones, VR technologies in public spaces. To begin with, from an analytical perspective, we suggest designers should not see the urban behavioural pattern of bodily passivity and social inattentiveness as a problem to solve or a flaw to fix. Nor shall researchers see the increasing bodily and communicative constraints brought by VR technology as a negative result. Based on this theoretical understanding, to form design opportunities, we suggest designers turning the 'civil inattention' towards resources for design and look at the effective social function. The focus could shift to the private space in people's minds and bring more joy and excitement away from encouraging people to be more bodily or socially active. However, we want to claim that we are not against designing for conviviality and social interactions in public spaces. Social connectedness and attachment are important elements for experiential aesthetics and wellbeing. More accurately, the learning we intend to convey is that designers should be more aware that when attempting to activate people with more interaction in the real social situation, such intervention might receive resistance as it can make the existing structure dysfunctional.

### 5.2 Strategies of VR content tailored for the metro context

The result of our study shows VR technology successfully brought more positive experiences, like making participants feel calmer and more relaxed, and less bored and stressed (Figure 3). We have described that the desirable experiences brought by the intervention were mainly associated with Video 3 (with four features discussed in section 'Result'). Now, we propose three strategies of 360° content that would be both acceptable and desired by metro passengers, which hopefully provides a larger implication on the application of VR in public transportations.

The first is the most straightforward, like other studies of mobile VR have suggested (Hock et al. 2017). The VR or 360° content shall play well with the moving status of wearers. If the scene also has speed, like the mobile setting of a car and a plane, or sporty activities like walking, running or swimming, the wearer will gain more sense of immersion.

Secondly, wearers shall feel they are within the scene as participants instead of watching the scene as an audience. This strategy mainly responds to the issue of continuous interest and immersion beyond novelty. P6 mentioned that her VR experience of watching an exotic beautiful scene was impressive. However, soon she got bored and lost interest. As implications, watching a scene is not adequate to gain constant attention. The virtual scene needs to give wearers a position or a role so it becomes his/her story. As one of the participants nicely framed '*in Video 1 I am alone; in Video 2 I am watching someone from a distance; in Video 3 I am with someone.*' The sense of participation of 'we three are going out on a boat for fishing' indicates the self-projection with a possibility in the virtual world. Therefore, the quality of the VR scene lies in whether it can facilitate the wearer projecting an imagined self instead of the marvellous and beautiful scene itself. The logic of the strategy shares insights from the concept of narcissus shoppers who seek the projection of an imagined or ideal self from the inspirational resources of urban space (Mäenpää 2005).

Thirdly, if the wearer needs to have a role in the VR scene, what kind of roles could it be? We argue the role shall be bodily constrained to be aligned with the behavioural norms in the metro. The VR content shall not encourage the wearer to become bodily active but, instead, support the acts of sitting and watching. Video 3 shows a very good example. The wearer was one of the three people on the boat, enjoying the moment but only sitting and watching. Just as P1 claimed '*it was*

*surprisingly relaxing to sit next to someone who is fishing than taking the metro*'. More specifically, the scene should be carefully designed in a way in which the wearer does not turn the head around too much as P5 pointed out that the most awkward act would be 'staring' at the passenger sitting next. For instance, the content we recommend could keep wearers focusing more or less on one direction, like sitting next to the person who is making pottery, fishing, cutting wood or climbing rocks.

In all, the implication we learn from this pilot study is that narrative-driven content works better rather than purely aesthetically pleasing scenes. If narratives are pedagogical, like pottery making or fishing, from which people can learn new skills and knowledge, they might be more motivated to watch continuously. Furthermore, proposed by P7, live streaming of a place somewhere on this planet, like a lake, bird's nest, or a busy square, would attract her with constant curiosity as well.

# 6 Conclusion

This study was interested in the broader application of VR technology in people's daily events and routines. We chose the mobile context of metro commuting in which we discussed the potential of VR technology in transforming commuting experiences. We looked at two perspectives of how VR technology was accepted socially and what VR content passengers were found desirable. The findings show that VR technology did not fundamentally challenge the norm of 'passive and anonymous in body, active in mind' in the Finnish metro. At the same time, VR technology successfully brought more positive experiences to metro rides. Moreover, this paper associates the enquiry of VR applications in the wild with the urban behavioural theories that provide a different view of 'a happy urban man'. As a contribution, it provides a theoretical lens to imply design practices of VR technology in problematising reality and suggesting new desirable possibilities.

Furthermore, we proposed three strategies of VR content that would be both appropriate and desirable in the context of the metro ride. Firstly, we recommend VR scenes could have movement and speed which synchronizes with the metro movement to provide a more sense of immersion. This can be applied to general travel contexts. Secondly, it is important for passengers to find a role of insiders within the scene rather than being an audience or outsiders watching the scene. And the method of engagement that is proved useful in our study is through narratives of activities, like '*I am one of three guys on this fishing trip*'. Thirdly, it is crucial to keep passengers bodily constrained and passive in order not to break the norms in metro cabins. Therefore, the fictional role in the VR scene could support the acts of sitting and watching just like what the metro passenger is supposed to do in reality.

This study has several limitations which require work for further exploration. Firstly, our study shows participants overall considered 'safe' to wear the VR headset in the metro. However, this result related to positive experiences and high social acceptance is difficult to be generalized because the Finnish metro is much emptier and safer compared with metro situations in other countries. In addition, the findings and takeaways are from a pilot study of one-time VR experiences. To discuss the potential of VR experiences integrating with daily commuting, the biggest challenge is to motivate passengers to pick up the headset again. Long-term intervention should be conducted in this matter.

## Acknowledgements

This research is funded by Aalto University Seed Funding. Special thanks to Johannes Söderström and Vreal for creating the three 360º videos for this research project.

# References

Ahmadpour, N., Kühne, M., Robert, J.-M. & Vink, P. 2016. Attitudes towards personal and shared space during the flight. *Work,* 54**,** 981-987.

Akiduki, H., Nishiike, S., Watanabe, H., Matsuoka, K., Kubo, T. & Takeda, N. 2003. Visual-vestibular conflict induced by virtual reality in humans. *Neuroscience letters,* 340**,** 197-200.

Augé, M. 1995. *Non-places: Introduction to an Anthropology of Supermodernity*, verso.

Chan, L. & Minamizawa, K. 2017. FrontFace: facilitating communication between HMD users and outsiders using front-facing-screen HMDs. *Proceedings of the 19th International Conference on Human-Computer Interaction with Mobile Devices and Services.*

Cresswell, T. 2006. *On the move: Mobility in the modern western world*, Taylor & Francis.

Davis, F. D., Bagozzi, R. P. & Warshaw, P. R. 1989. User acceptance of computer technology: A comparison of two theoretical models. *Management science,* 35**,** 982-1003.

Edensor, T. 2011. Commuter: mobility, rhythm and commuting. *Geographies of mobilities: Practices, spaces, subjects***,** 189-204.

Eghbali, P., Väänänen, K. & Jokela, T. Social acceptability of virtual reality in public spaces: experiential factors and design recommendations.  Proceedings of the 18th International Conference on Mobile and Ubiquitous Multimedia, 2019. 1-11.

Goffman, E. 2005. *Interaction ritual: Essays in face to face behavior*, AldineTransaction.

Groening, S. J. F. H. 2016. "No One Likes to Be a Captive Audience": Headphones and In-Flight Cinema. 28**,** 114-138.

Gugenheimer, J. 2016. Nomadic virtual reality: Exploring new interaction concepts for mobile virtual reality head-mounted displays. *Proceedings of the 29th Annual Symposium on User Interface Software and Technology.*

Gugenheimer, J., Dobbelstein, D., Winkler, C., Haas, G. & Rukzio, E. Facetouch: Enabling touch interaction in display fixed uis for mobile virtual reality.  Proceedings of the 29th Annual Symposium on User Interface Software and Technology, 2016. 49-60.

Gugenheimer, J., Mai, C., Mcgill, M., Williamson, J., Steinicke, F. & Perlin, K. 2019. Challenges using head-mounted displays in shared and social spaces. *Extended Abstracts of the 2019 CHI Conference on Human Factors in Computing Systems.*

Gugenheimer, J., Stemasov, E., Sareen, H. & Rukzio, E. 2017. FaceDisplay: enabling multi-user interaction for mobile virtual reality. *Proceedings of the 2017 CHI Conference Extended Abstracts on Human Factors in Computing Systems.*

Halse, J. & Boffi, L. 2016. Design interventions as a form of enquiry. *In:* SMITH, R. C., VAN HEESWIJK, J., KJÆRSGAARD, M., OTTO, T., HALSE, J. & BINDER, T. (eds.) *Design anthropological futures.* London: Bloomsbury Publishing.

Hirzle, T., Rixen, J., Gugenheimer, J. & Rukzio, E. Watchvr: Exploring the usage of a smartwatch for interaction in mobile virtual reality.  Extended Abstracts of the 2018 CHI Conference on Human Factors in Computing Systems, 2018. 1-6.

Hock, P., Benedikter, S., Gugenheimer, J. & Rukzio, E. 2017. Carvr: Enabling in-car virtual reality entertainment. *Proceedings of the 2017 CHI Conference on Human Factors in Computing Systems.*

Hutchinson, H., Mackay, W., Westerlund, B., Bederson, B. B., Druin, A., Plaisant, C., Beaudouin-Lafon, M., Conversy, S., Evans, H. & Hansen, H. Technology probes: inspiring design for and with families. Proceedings of the SIGCHI conference on Human factors in computing systems, 2003. 17-24.

Kauhanen, O., Väätäjä, H., Turunen, M., Keskinen, T., Sirkkunen, E., Uskali, T., Lindqvist, V., Kelling, C. & Karhu, J. 2017. Assisting immersive virtual reality development with user experience design approach. *Proceedings of the 21st International Academic Mindtrek Conference.*

Kelling, C., Väätäjä, H. & Kauhanen, O. 2017. Impact of device, context of use, and content on viewing experience of 360-degree tourism video. *Proceedings of the 16th International Conference on Mobile and Ubiquitous Multimedia.*

Kola, J.-P. 2011. *Spectacular Place,* Helsinki, Aalto University.

Kuznetsov, S. & Paulos, E. Participatory sensing in public spaces: activating urban surfaces with sensor probes.  Proceedings of the 8th ACM Conference on Designing Interactive Systems, 2010. 21-30.
Lefebvre, H. & Nicholson-Smith, D. 1991. *The production of space,* Oxford, UK, Blackwell Publishing.
Lyons, G., Jain, J. & Holley, D. 2007. The use of travel time by rail passengers in Great Britain. *Transportation Research Part A: Policy,* 41**,** 107-120.
Mäenpää, P. 2005. *Narkissos kaupungissa: tutkimus kuluttaja-kaupunkilaisesta ja julkisesta tilasta,* Helsinki, University of Helsinki.
Mäenpää, P. 2012. Spicing up metro stations–urban sociological approach. *Understanding Public Spaces Through Narrative Concept Design.* Helsinki: School of Art and Design, Aalto University.
Mai, C. & Khamis, M. 2018. Public HMDs: Modeling and Understanding User Behavior around Public Head-Mounted Displays. *Proceedings of the 7th ACM International Symposium on Pervasive Displays.*
Mai, C., Rambold, L. & Khamis, M. 2017. TransparentHMD: revealing the HMD user's face to bystanders. *Proceedings of the 16th International Conference on Mobile and Ubiquitous Multimedia.*
Mattelmäki, T. 2006. *Design probes,* Helsinki, UIAH.
Mattelmäki, T. & Vaajakallio, K. (eds.) 2012. *Understanding Public Spaces Through Narrative Concept Design,* Helsinki: School of Art and Design, Aalto University.
Mcgill, M., Boland, D., Murray-Smith, R. & Brewster, S. 2015. A dose of reality: Overcoming usability challenges in vr head-mounted displays. *Proceedings of the 33rd Annual ACM Conference on Human Factors in Computing Systems.*
Mcgill, M. & Brewster, S. Virtual reality passenger experiences.  Proceedings of the 11th International Conference on Automotive User Interfaces and Interactive Vehicular Applications: Adjunct Proceedings, 2019. 434-441.
Mcgill, M., Ng, A. & Brewster, S. I am the passenger: how visual motion cues can influence sickness for in-car VR.  Proceedings of the 2017 chi conference on human factors in computing systems, 2017. 5655-5668.
Mcgill, M., Williamson, J., Ng, A., Pollick, F. & Brewster, S. 2019. Challenges in passenger use of mixed reality headsets in cars and other transportation. *Virtual Reality***,** 1-21.
Negishi, K. & Bissell, D. 2020. Transport imaginations: Passenger experiences between freedom and constraint. *Journal of Transport Geography,* 82**,** 102571.
O'hagan, J. & Williamson, J. R. Reality aware VR headsets.  Proceedings of the 9TH ACM International Symposium on Pervasive Displays, 2020. 9-17.
Paredes, P. E., Balters, S., Qian, K., Murnane, E. L., Ordóñez, F., Ju, W. & Landay, J. A. 2018. Driving with the Fishes: Towards Calming and Mindful Virtual Reality Experiences for the Car. *Proceedings of the ACM on Interactive, Mobile, Wearable Ubiquitous Technologies.*
Rahimizhian, S., Ozturen, A. & Ilkan, M. 2020. Emerging realm of 360-degree technology to promote tourism destination. *Technology in Society,* 63**,** 101411.
Remesar, A. 2021. Co-design of public spaces with local communities. *The Palgrave Handbook of Co-Production of Public Services and Outcomes.* Springer.
Schwind, V., Reinhardt, J., Rzayev, R., Henze, N. & Wolf, K. 2018. Virtual reality on the go?: a study on social acceptance of VR glasses. *Proceedings of the 20th International Conference on Human-Computer Interaction with Mobile Devices and Services Adjunct.* Barcelona, Spain: ACM.
Seeburger, J. & Foth, M. Content sharing on public screens: experiences through iterating social and spatial contexts.  Proceedings of the 24th Australian Computer-Human Interaction Conference, 2012. 530-539.
Sennett, R. 2017. *The fall of public man,* London, WW Norton & Company.
Sholette, G. & Thompson, N. 2004. The interventionists: user's manual for the creative disruption of everyday life. *North Adams: MASS MoCA*.
Simmel, G. 2012. The metropolis and mental life. *The urban sociology reader.* Routledge.

Smus, B. & Riederer, C. Magnetic input for mobile virtual reality. Proceedings of the 2015 ACM International Symposium on Wearable Computers, 2015. 43-44.
Turner, W. R., Nakamura, T. & Dinetti, M. 2004. Global urbanization and the separation of humans from nature. *Bioscience,* 54**,** 585-590.
Vroman, L. & Lagrange, T. 2017. Human movement in Public spaces: The use and development of motion-oriented design strategies. *The Design Journal,* 20**,** S3252-S3261.
Williamson, J. R., Mcgill, M. & Outram, K. 2019. Planevr: social acceptability of virtual reality for aeroplane passengers. *Proceedings of the 2019 CHI Conference on Human Factors in Computing Systems.*
Wu, Y. 2017. *Bicycles and plants: Designing for conviviality and meaningful social relations through collaborative services,* Helsinki, Aalto University.
Yang, K.-T., Wang, C.-H. & Chan, L. Sharespace: Facilitating shared use of the physical space by both vr head-mounted display and external users. Proceedings of the 31st Annual ACM Symposium on User Interface Software and Technology, 2018. 499-509.
Yang, T., Lai, I. K. W., Fan, Z. B. & Mo, Q. M. 2021. The impact of a 360° virtual tour on the reduction of psychological stress caused by COVID-19. *Technology in Society,* 64**,** 101514.